\documentclass[12pt]{article}
\input epsf.tex
\def\be{\begin{equation}}
\def\ee{\end{equation}}
\def\bea{\begin{eqnarray}}
\def\eea{\end{eqnarray}}
\def\tr{{\rm tr}}
\def\dd{\partial}
\def\la{\raise.16ex\hbox{$\langle$}\lower.16ex\hbox{}  }
\def\ra{\, \raise.16ex\hbox{$\rangle$}\lower.16ex\hbox{} }
\def\go{\rightarrow}

\def\next{{~~~,~~~}}
\def\onehalf{ \hbox{${1\over 2}$} }

\def\ep{\epsilon}
\def\myfrac#1#2{{\mybig #1 \over \mybig #2}}

\def\mypagenumber{1}
\def\mydate{\today}
\def\myend{\end{document}}

\def\Journal#1#2#3#4{{#1}{\bf #2} (#4) #3}

\def\NPB{{\em Nucl.\ Phys.} B}
\def\PLB{{\em Phys.\ Lett.} B}
\def\PRL{\em Phys.\ Rev.\ Lett. }

\def\PRD{{\em Phys.\ Rev.} D}
\def\AP{{\em Ann.\ Phys.\ (N.Y.)} }

\normalsize
\let\oldtheequation=\theequation
\def\doteqs#1{\setcounter{equation}{0}
            \def\theequation{{#1}.\oldtheequation}}
\newcounter{sxn}
\def\sx#1{\addtocounter{sxn}{1} \vskip 1.cm  \goodbreak
\noindent{\large\bf\leftline{\thesxn.~~#1}} \nobreak \vskip -.5cm}
\def\sxn#1{\sx{#1} \doteqs{\thesxn}}

\newcounter{axn}

\date{}

\newdimen\mybaselineskip
\newcommand{\bpic}{\begin{picture}}
\newcommand{\epic}{\end{picture}}

\def\mybig{\displaystyle \strut }

\def\dd{\partial}
\def\la{\raise.16ex\hbox{$\langle$} \, }
\def\ra{\, \raise.16ex\hbox{$\rangle$} }
\def\go{\rightarrow}

\def\next{{~~~,~~~}}
\def\ie{{\it i.e., \ }}

\def\onehalf{ \hbox{${1\over 2}$} }

\def\psibar{ \psi \kern-.65em\raise.6em\hbox{$-$} }
\def\mbar{ m \kern-.78em\raise.4em\hbox{$-$}\lower.4em\hbox{} }

\def\phibar{{\bar\phi}}

\def\ep{\epsilon}

\def\myfrac#1#2{{\mybig #1\over \mybig #2}}

\def\n@space{\nulldelimiterspace=0pt \mathsurround=0pt }
\def\huge#1{{\hbox{$\left#1\vbox to 20.5pt{}\right.\n@space$}}}

\def\myskip{\noalign{\kern 8pt}}
\def\myeqspace{\noalign{\kern 10pt}}

\def\boxit#1{$\vcenter{\hrule\hbox{\vrule\kern3pt
    \vbox{\kern3pt\hbox{#1}\kern3pt}\kern3pt\vrule}\hrule}$}
\def\bigbox#1{$\vcenter{\hrule\hbox{\vrule\kern5pt
     \vbox{\kern5pt\hbox{#1}\kern5pt}\kern5pt\vrule}\hrule}$}

\def\ignore#1{{}}

\begin{document}
\bibliographystyle{unsrt}
\footskip 1.0cm

\thispagestyle{empty}
\setcounter{page}{\mypagenumber}

%{\baselineskip=10pt \parindent=0pt \small
%\mydate 
%}                             
\begin{flushright}{\mydate \\  UMN-TH-11712/98  \\ TPI-MINN-98/12-T \\
hep-th/9808057}
\\
\end{flushright}

\vspace{0.3cm}
\begin{center}
{\large \bf {Complex  Monopoles in YM + Chern-Simons }}\\
\vskip .4cm 
{\large \bf {Theory in 3 Dimensions}\footnote{Talk given by K. Saririan at the Quarks-98 Conference in Suzdal, Russia (May 1998). This work was supported in part  by the U.S. Department of Energy under contracts DE-FG02-94ER-40823.}
}\\
\vspace{0.6cm}
{ Bayram Tekin, Kamran Saririan and Yutaka Hosotani}

\vspace{.5cm}
{\it School of Physics and Astronomy, University of Minnesota}\\ 
{\it  Minneapolis, MN 55455, U.S.A.}\\ 
\end{center}

\vspace*{0.5cm}
%\baselinestretch{2.0}

%\normalsize

\begin{abstract}
In this talk, we give a brief discussion of  complex monopole solutions in the  three dimensional 
Georgi-Glashow model with a Chern-Simons term.  
We find that there exist complex monopole
solutions of finite action.  They dominate the path integral and 
disorder the Higgs vacuum, but electric charges are not confined.
Subtleties in the model and  
issues related to Gribov copies are also noted.
\end{abstract}

\sxn{Introduction}

Monopole solutions can exist in gauge theories in which there is an unbroken compact U(1)group. The three dimensional SO(3) Yang-Mills theory with an adjoint Higgs field which breaks the $SO(3)$ to $U(1)$ (also known as the Georgi-Glashow model) was shown by Polyakov \cite{Polyakov} to have monopole solutions which lead to the linear confinement of electric charges. The monopole contribution causes the Higgs vacuum to be `disordered', \ie  
\( \langle h^a \rangle = 0\) while \( \langle h_ah^a\rangle=v^2\ne0 . \)
In other words, long-range order in the Higgs vacuum is destroyed by the existence of monopoles. If now one adds the Chern-Simons (CS) term, it can be readily seen that \cite{Deser}  the gauge fields acquire topological mass.
the unbroken $U(1)$ gauge field also becomes massive.
As a consequence, the linear confinement disappears in the presence of the CS
term:  there is no long-range force in the Georgi-Glashow-Chern-Simons 
(GGCS) model to start with.  The electric
flux is not conserved.  It does not matter for the issue of the 
confinement whether or not monopole configurations dominate in the 
functional integral. 
 
Recall that the action for the GGCS model is given by:
\be S = S_{\rm YM} + S_{\rm CS} + S_{\rm H}  \hspace{0.3in} {\rm (Euclidean)}\ee
where
\bea
S_{\rm YM} & = & - \myfrac{1}{2g^2} \int d^3x\; \tr \, F^2  \nonumber \\
S_{\rm CS} & = & - \myfrac{i \kappa}{g^2} \int d^3x \ep^{\mu\nu\lambda}\; \tr (A_{\mu}\dd_\nu A_{\lambda} + \myfrac{2}{3} A_{\mu} A_{\nu} A_{\lambda} )
 \nonumber \\
S_{\rm H} & = & \myfrac{1}{g^2} \int d^3x [ \onehalf \left(D_{\mu}h^a\right)^2 + \myfrac{\lambda}{4} \left(h_ah^a - v^2\right)^2 ] \nonumber 
\eea
Notice that  $S_{\rm SC}$ is pure imaginary, \ie the action is complex. Therefore it is natural to seek complex solutions that extremize the complex action. 
In the following, we shall explore the complex monopole solutions and address the following issues:

\noindent
$\bullet$
 Are there finite action complex monopole solutions? \\
$\bullet$
See the connection with gauge (Gribov) copies (i.e., do they cancel or wipe out the effect of monopoles, as in the real monopole case?
\\
$\bullet$
 What happens to the long-range order in Higgs vacuum? 

\noindent
We shall address these questions within a saddle point approximation.   
Few details are presented below, but they could be found in ref. \cite{us}. 
However, before discussing the complex monopoles, it is worthwhile to recall a few facts of life about the model and the scenario of the real monopole solutions in this model. Many authors have shown that there is no real
monopole-type field
configuration of a finite action which solves the equations of motion \cite{Pisarski}-\cite{Fradkin}.
This fact
has been interpreted as indicating the irrelevance of (real) monopole
configurations in the model. 

\vspace{0.2in}

\noindent 
{\sc The Model}

\vspace{0.15in}

The standard monopole ansatz   \cite{tHooft} for the action given by eq. (1) is given by: 
\bea
&&h^a(\vec{x})=\hat x^a h(r) \cr
&&A_{a\mu}(\vec{x})= {1\over r} \left[ \epsilon_{a\mu
\nu}\hat{x}_\nu(1-
\phi_1
)+ (\delta_{a\mu}-\hat{x}_a 
\hat{x}_\mu) 
\phi_2 
+ r 
A 
\hat{x}_a \hat{x}_\mu \right]     
\eea  
with boundary conditions: 
\bea
&&h=0 \next  \phi_1 = 1 \next  \phi_2 =0 \quad {\rm at} ~ r=0 \cr
&& h=v \next \phi_1 =0 \next \phi_2 = 0 \quad {\rm at} ~ r=\infty  \nonumber 
\eea
Recall that with this ansatz the 't Hooft-Polyakov monopole has 
\( h, \phi_1 \ne 0\), and \(\phi_2=A=0. \)
 The residual $U(1)$ symmetry in the model after the breaking $SO(3)\rightarrow U(1)$ is parameterized using again the standard abelian projection as follows: \be
\Omega = \exp \left(\myfrac{i}{2} f(r) \hat x_a\tau^a\right) .
\ee
Under this symmetry the components of the gauge field (2) transform as:,
\bea  
\left( \begin{array}{c} \phi_1 \\ \phi_2 \end{array} \right) & \go &
\left( \begin{array}{cc} 
 \cos f & \sin f \\ 
-\sin f  & \cos f \end{array} \right) 
\left( \begin{array}{c} \phi_1 \\ \phi_2 \end{array} \right)
\cr & &  \cr 
A  & \go &  A - f' .\nonumber 
\eea
The equations of motion are invariant under this transformation but the
action (1.1) is not. 
Under a more general gauge transformation $A \go \Omega A \Omega^{-1}
+ g^{-1} \Omega d \Omega^{-1}$, 
\be
\delta_g S_{CS} = {i\kappa\over g^2} \int 
   {\rm tr~} d(A \wedge d\Omega^{-1} \Omega)
+ {i\kappa\over 3g^2} \int {\rm tr~} d\Omega^{-1} \Omega
   \wedge d\Omega^{-1} \Omega \wedge d\Omega^{-1} \Omega ~~.
\label{GT1}
\ee
If the theory is defined on $S^3$, the first term vanishes.
The second term is the winding number of the mapping $\Omega$.
This leads to the quantization condition
$4\pi\kappa/g^2 = n =$ an integer \cite{Deser}.

On $R^3$, however, the first term does not vanish for monopole
configurations.  Under the transformation (1.3) the first and last terms
on the r.h.s. of eq. (1.4) are $(4\pi i \kappa/g^2) \sin f(\infty)$ and
$(4\pi i \kappa/g^2) ( f(\infty) -\sin f(\infty))$, respectively.
Hence
\be 
 S_{CS}\go S_{CS} + {4\pi i\kappa\over g^2}  f(\infty)  ~~~.
\label{transform2}
\ee
This raises a puzzle regarding the quantization of this theory. The usual Faddeew-Popov procedure   
\[ {\cal Z} \sim \int[{\cal D}A_\mu][{\cal D}h]\Delta_F(A_\mu)\delta[{\cal F}(A_\mu)]e^{-S[A_\mu, h]}
\]
works fine if   $\delta_g S = 0$, but not for a non-invariant action. For a more detailed discussion (but not a resolution!) of this puzzle, see ref. \cite{us}. Let us just remark that at  this stage, to our knowledge, there is no consistent formulation of this model beyond semi-classical level.

\vspace{0.2in} 

\noindent
{\sc (Real) Monopole Solutions}

\vspace{0.15in}

As mentioned earlier, the real monopole solutions of the GGCS model have been the subject of several previous studies, with the similar conclusion that the
contribution  of real monopoles in this model vanishes in the partition function. Pisarski \cite{Pisarski} showed that real monopoles give rise to a linearly divergent action, which makes (real) monopoles irrelevant, and furthermore, the linearly divergent action leads to the  interpretation that the monopole-antimonopole pairs are confined. In ref. \cite{Affleck}, the authors reach a similar conclusion. There, in the gauge transformation of the full action, what we have referred to as $f(\infty)$ in: 
$ S_{\rm CS} \rightarrow S_{\rm CS} + i n f(\infty) $ is interpreted as the collective coordinate associated with the monopole solution, and is restricted to a $2\pi$ interval. They argue that, therefore, integrating over this collective coordinate\footnote{We shall instead interpret $f(\infty)$ as the parameter associated with the gauge copies (or Gribov copies) which is also summed over in the partition function.} wipes out the  monopole contribution in the path integral:  

\[ {\cal Z} \propto \int df(\infty) e^{i n f(\infty)} = 0 . \]

We will re-examine these considerations for complex monopoles shortly. We do this by considering the complex monopole solutions for different gauge choices.

\sxn{ `Axial' gauge: $\hat x_\mu A_\mu = 0$}

In this gauge,  the component $A=0$ (see eq. (1.2)) and the remaining component fields  satisfy the following equations of motion:

\be
\cases{
\left( \myfrac{d^2}{dr^2} - \myfrac{1}{r^r}(\phi_1^2 + \phi_2^2 - 1) - h^2\right) 
\left( \begin{array}{c} \phi_1 \\ \phi_2 \end{array} \right) 
+ i \kappa \myfrac{d}{dr} \left( \begin{array}{c} \phi_2 \\ -\phi_1 \end{array} \right) = 0  
\cr 
h'' + \myfrac{2}{r} h' - {\lambda(h^2 - v^2) + \myfrac{2}{r} (\phi_1^2 + \phi_2^2)} h = 0 . }  \ee
 In this gauge, complex solutions exist, which satisfy:
\[ \myfrac{\phi_2}{\phi_1} = i \tanh \left(\myfrac{\kappa r}{2}\right) \]
 where $\kappa = \myfrac{n g^2}{4 \pi}$ and boundary conditions:
\[ 
\cases{
\phi_1 (0) = 1,   \hspace{2cm} \phi_1(\infty)=0    \cr
  \phi_2 (0) = 0,   \hspace{2cm} \phi_2(\infty)=0 }
\]
Let us simply outline how the solutions in th axial gauge are obtained. First we choose the coordinates: 
\bea
\eta  &=& (\phi_1 + i \phi_2)e^{- i\int^r A(r) dr}  \cr
\zeta   &=& (\phi_1 - i \phi_2)e^{+ i\int^r A(r) dr}  \nonumber 
\eea
in which the action is written as follows:
\bea
S &=& {4\pi\over g^2} \int_{0}^{\infty} dr [ \zeta'\eta'+ 
    {1\over 2 r^2}(1- \eta\zeta)^2 - {\kappa\over 2} (\eta'\zeta 
    -\eta \zeta') \cr
 &+& {r^2\over 2} (h')^2 + h^2 \eta \zeta + 
{\lambda r^2\over 4}(h^2-v^2)^2  + i \kappa A ]
. \eea
We then let $A=0$ (\ie fixing the axial gauge). Notice that the action in this gauge becomes real.  The BC's and the equations of motion result in the relation: \( \zeta = \eta \exp{\kappa r}\). 
In the limit of no  CS term ($\kappa=0 \hspace{1cm} \Rightarrow \zeta=\eta$) the equations of motion reduce to:
\[
\cases{
\zeta'' + \myfrac{\zeta}{r^2}(1 - \zeta^2) - h^2 \zeta = 0 \cr
(r^2 h')' - (2 \zeta^2 - \mu^2 r^2 + \lambda r^2 h^2) h  = 0 
} 
\]
In the limit $\lambda=0$, the exact solutions for these equations are the well known BPS solutions:
\be \cases{
h(r)= v \coth(v r) -\myfrac{1}{r}  \cr
\zeta(r) = 
\myfrac{v r}{ \sinh(v r) }
}
\ee
We use these solutions, and numerically generate the solutions 
for $\lambda\ne 0$ and $\kappa \ne 0$. These solutions are shown in Fig. 1,
These complex solutions have the following features which 
are characteristic of monopoles: They minimize  the 
\underline{real} action (in axial gauge), they
 have mass  $\propto 1/g^2$ and the  $U(1)$ field strength:
\( F_{\mu\nu}= -\myfrac{1}{r^2}\hat x_a \epsilon_{a\mu\nu} .\) 
Notice also that in this gauge there are no Gribov copies. 

%%%%%%%%% Insert a plot of solutions in axial gauge %%%%%%%%%%%%
\begin{figure}[ht]\centering
% \leavevmode 
\mbox{
\epsfysize=2.2in \epsfbox{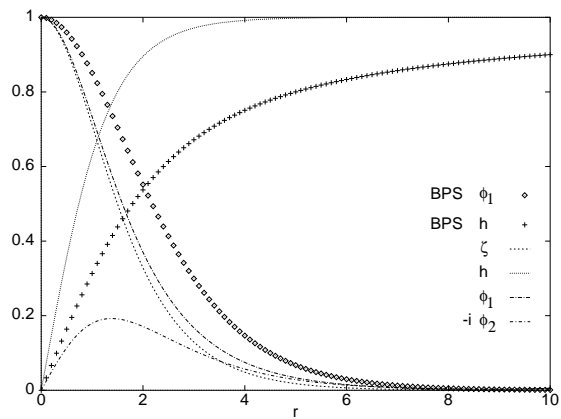}}
\caption{Axial gauge: $\myfrac{\kappa}{2v}=0.25$ and $\lambda=0.5$.} 
\end{figure}

\sxn{Radiation gauge ($D_\mu A_\mu^a=0$)}

In terms of the ansatz (eq. (1.2)), but again with $\phi_1, \; \phi_2, \;$ and $A$ complex,  the gauge condition gives rise to a relations between $ \phi_2$ and $A$, namely:
\be A' + \myfrac{2}{r} A -  \myfrac{2}{r^2} \phi_2 = 0 \;\; {\rm or} \;\;
 A(r) = {2 \over r^2} \int_0^r dr \phi_2(r) \ee 
plus the equations of motion for $\phi_1$, $\phi_2$, and $h$, (resp.)  in this gauge: 
\bea
&& (\phi_1' +A \phi_2)' +A(\phi_2'-A\phi_1) + \myfrac{1}{r^2}
(1-\phi_1^2 -\phi_2^2)\phi_1 
 +i\kappa (\phi_2'- A\phi_1) -h^2\phi_1 = 0
, \\
 && (\phi_2' -A \phi_1)'- A(\phi_1'+A\phi_2 ) + \myfrac{1}{r^2}
(1-\phi_1^2 -\phi_2^2)\phi_2   -  i\kappa (\phi_1'+ A\phi_2)  
-h^2\phi_2 \cr 
&& = 2\int_r^\infty du  \myfrac{1}{u^2}[\phi_2\phi_1' -  \phi_1 \phi_2' +A(\phi_1^2 + \phi_2^2) + \myfrac{i\kappa}{2}(\phi_1^2 + \phi_2^2 - 1)] ,
\\
&& (r^2h')' - \lambda r^2(h^2 + v^2)h - 2 (\phi_1^2 + \phi_2^2)h = 0 . \eea
Exact numerical solutions of these equations for $\phi_1$ and $\phi_2$ is rather difficult, and will be determined in the future, but our preliminary numerical analysis indicates that the solutions exist. For the purposes of studying the question of gauge copies which we turn to next, we  use lump-like ansatz for these solutions that are consistent with the asymptotic equations.

 Recall the residual $U(1)$  with:
\(
\Omega = \exp(\myfrac{i}{2} f(r) \hat x_a\tau^a)
 \). In this gauge however, there is a nontrivial condition on  
$f(r)$ \cite{Gribov}: 
\be
f''  + \myfrac{2}{r}f' - \myfrac{2}{r^2} \left( \phi_1 \sin f
+ \phi_2 (1-\cos f) \right) = 0 .
\ee
Also recall the non-invariance of the CS term:
\[ S_{CS} \longrightarrow S_{CS} + i n f(\infty) . \]   
We now ask the following question:  in real 
monopole case, summing over the gauge copies (or alternatively integrating 
over the collective coordinate of the monopole \cite{Affleck}) leads to 
cancellation. Is this true for the complex monopoles as well? In other words, 
\[ 
\displaystyle\sum_{\rm Gribov\;\; copies} e^{i n f(\infty)}\; \stackrel{??}{=}\; 0 . \]

\vspace{0.2in}

\noindent
{\sc Gribov Copies}

\vspace{0.2in}

We now examine  the question of Gribov copies by looking at the solutions $f(\infty)$ of the Gribov equation (3.5) with the appropriate initial conditions at $r=0$. First we consider the trivial case of vacuum solutions 
($A_\mu^a=0$)
 i.e.,  $\phi_1=1$, $\phi_2=0$:
\[ f'' + {2 \over r} f' - {2 \over r^2} \sin f = 0 . \]
This implies the following solutions:
\[ 
f(0)=0 \Longrightarrow \cases{(i)\;\; f(\infty)=0 \hspace{1cm} 
\leftrightarrow \; \; A_\mu^a=0 \cr
(ii),(iii)\;\;  f(\infty)= \pm \pi \hspace{1cm} 
\leftrightarrow \; \;  {\rm Gribov\;\; Copies}  }
\]

One of the Gribov copies ($+\pi$) is shown in Fig. 2. For various (positive) $f'(0)$, the value of $f(\infty)$ approached is unique -- no surprise in the case of the vacuum solution. 

  \begin{figure}[ht]\centering
 \leavevmode 
\mbox{
\epsfysize=2in \epsfbox{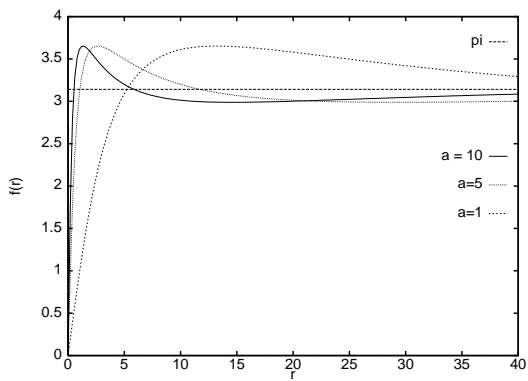}}
\caption{$f(r)$ for vacuum configuration. There are similar curves (not shown) approaching $-\pi$ for negative initial values of $f'(0)$.} 
\end{figure}

It is interesting to consider the 
't Hooft-Polyakov monopole $\phi_2=0$ 
but $\phi_1\not= 0$ ($\phi_1(0)=1$ and $\phi_1 \go 0$ as 
$r\go \infty$).\footnote{
 Although this is clearly not a solution to the theory including
the CS term, this exercise illustrates the possibility that summing of 
the Gribov copies may not lead to the cancellation of the monopole-type 
contribution.}
  In terms of the Gribov equation, this case corresponds to a 
particle moving 
in a time-dependent potential.  As the potential  becomes exponentially
small for large $t$, there can be a continuous family of solutions
parameterized by the value of $f(\infty)$.   The asymptotic value
$|f(\infty)|$ depends of the initial slope $f'(0)$.  In the BPS limit it 
ranges from $0$ to $3.98$ (see Fig. 3).  For $|f'(0)| \ll 1$,
$|f(r)|$  remains
small.  For $|f(r)| \gg 1$, $f(r)$ approaches an asymptotic value
before $\phi_1$ and $\phi_2$ make sizable changes, {\it i.e., } 
$f(r)$ behaves as in the vacuum case.  The maximum value for
$|f(\infty)|$ is attained for $f'(0)=\pm 2.62$.

%%%%%%%% Insert plot for BPS
\begin{figure}[ht]\centering
 \leavevmode 
\mbox{
\epsfysize=2in \epsfbox{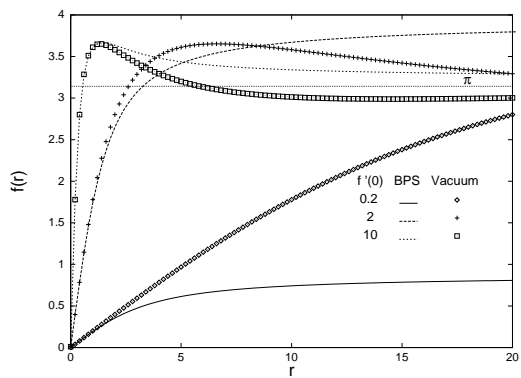}}
\caption{$f(r)$ for BPS monopole configuration. Solid lines correspond to monopoles (generally $\not\leq \pi$), other  lines are the vacuum solutions (all converging to $ \pi$ for positive initial slopes, $a$).} 
\end{figure}

We see that even for the BPS solutions, The values of $f(\infty)$ are \underline{not} restricted to $[-\pi, \pi]$, nor are they completely arbitrary. The plot in Fig. 4 shows the  range of $f(\infty)$ as a function of $f'(0)$ with the latter varying over five orders of magnitude.  
What we find is that $ 0\le f(\infty) < 3.98$, and so integration over all possible values of $ f(\infty)$ is not expected to cancel 
(cf. ref. \cite{Affleck}).

\begin{figure}[ht]\centering
 \leavevmode 
\mbox{
\epsfysize=2in \epsfbox{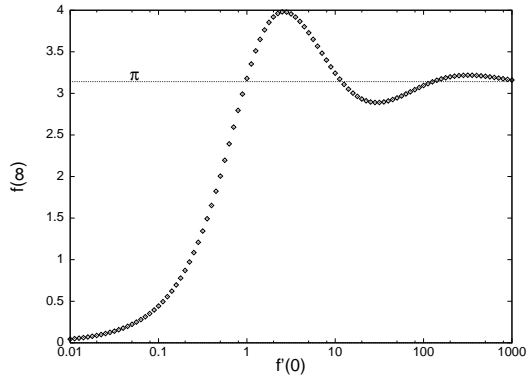}}
\caption{$f(\infty)$ for BPS monopole vs. initial slope ($f'(0)$).}
\end{figure}

Now let us consider the full theory with the Chern-Simons term. For complex $\phi_1$ and $\phi_2$, solutions of eq. (3.5) 
are necessarily complex. The lump-like ansatz that are used are such that 
$\phi_1$ is real and   $\phi_2$ is pure imaginary. 
Therefore $A$ is also pure imaginary (see eq. (3.1))
 We explore ``complex'' Gribov copies of the solution.
 There is no ``real'' Gribov copy.
With $f=f_R + i f_I$ eq. (3.5) becomes
\bea
&&f_R'' + {2\over r} f_R' 
-{2\over r^2} \Big\{ \phi_1 \sin f_R \cosh f_I
- \phibar_2 \sin f_R \sinh f_I \Big\} = 0 \cr
\noalign{\kern 10pt}
&&f_I'' + {2\over r} f_I' 
-{2\over r^2} \Big\{ \phi_1 \cos f_R \sinh f_I
+ \phibar_2 (1 - \cos f_R \cosh f_I) \Big\} = 0 ~.
\label{radiation4}
\eea
Boundary conditions are $f_R(0) = f_I(0)=0$ and 
$f_R(\infty), f_I(\infty)=$~finite.  Although the meaning of these
solutions is not clear for $f_I(r)\not= 0$, we point out that solutions
satisfying $f_I(\infty)=0$ might have special role in the path
integral.  In view of (1.4), such copies carry
the additional oscillatory factor $(4\pi i \kappa/g^2) f_R(\infty)$
in the path integral.  Examination of Eq.\ (\ref{radiation4}) with
representative $\phi_1$ and $\phibar_2$ shows that such a solution
is uniquely found with given $f_R'(0)$.  $f_R(\infty)$ is determined 
as a function of $f_R'(0)$.  The range of $f_R(\infty)$ is not
restricted to $[-\pi, \pi]$ etc.  We expect no cancellation
in the path integral from these copies. 
Therefore the monopole contribution to the 
path integral does not appear to be destroyed. 

\begin{figure}[ht]\centering
 \leavevmode 
\mbox{
\epsfysize=2.1in \epsfbox{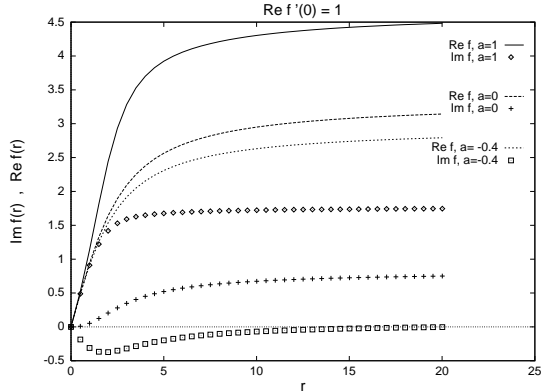}}
\caption{Complex $f(r)$ - YM-Chern-Simons.} 
\end{figure}
%%%%%%%%%%%%%  Plot for the CS term (real/pure imag ansatz)

\sxn{Conclusions and Discussion}

We have seen that in the GGCS theory, complex monopoles exist, and that they
have a non-vanishing contribution to the  path integral. As we have
shown,  the cleanest way to see this is in the radial gauge.
The action is  minimized by complex solutions, and is real and finite.
Furthermore, the solutions 
 have the usual  characteristics of monopoles.  They 
have $U(1)$ field strength 
given by $F_{\mu\nu}= -\epsilon_{\mu\nu}\hat x^a / r^2$ and  
mass $\sim 1/g^2$.  As a consequence, the 
long range order in the Higgs vacuum is destroyed. However, 
we must recognize that we are far from understanding this theory
at quantum level, or beyond semi-classical approximation. 
The understanding of the quantum theory is obscured by the 
gauge non-invariance of the CS term. 
 There are other murky areas in this model some of which are 
related to the puzzles about the gauge invariance and quantization.

We started with a theory with compact U(1) symmetry, where by definition  
the gauge transformation parameter (which we called $f$) is originally 
a real valued
function. However, in discussing the Gribov copy problem in the
radiation gauge, we have looked for complex field configurations
which are related to each other by complex $f$.

Curiously enough, the gauge invariant 
part of the action (\ie everything other than the 
CS term) is still invariant under the transformation with complex
$f$.   There are many saddle points in the complex field configuration 
space which are related by these complex $f$'s.
Nevertheless, the physical interpretation of $f$ being complex is not 
obvious at all. 

If we had restricted
the gauge parameter $f$ in this theory to be
real and had not allowed complex $f$, then the absence of real solutions to
the  Gribov  equation in the radiation gauge would have led to 
no Gribov copies. 
We have adopted the view that  complex solutions to Gribov's equation 
correspond to generalized Gribov copies of  complex saddle points. 
We understand that this is a question not completely settled, and
warrants further investigation. Meanwhile we conclude that within
our semi-classical approximation, it appears that  summation over 
the Gribov copies
(or integrating over the collective coordinates) of the complex 
monopole solutions does not lead to the cancellation of the monopole 
contribution.  What if quantum
corrections  to the Jacobian of the Gribov copies somehow cancel the
effect after all?   This is one of the questions which we could not answer in
this model unless we learn how to  go beyond the semi-classical approximation.
We remark that
one could raise the same objection for the real monopole case where 
it has been argued in the literature that 
the integral cancels.   Recall that the issues of the non-invariance 
of the CS
term,  and the problems associated with the quantization are
irrespective of whether the  monopole is real or complex.

Let us remark that complex gauge field configurations have been studied
before in the literature. In particular Wu and Yang \cite{Wu} have given a
prescription of how complex gauge fields in $SU(2)$ theory can be
converted to real gauge fields for the group $SL(2,C)$. Witten
\cite{Witten} shows that Chern-Simons theory with the group $SL(2,C)$
is equivalent to the 2+1 dimensional Einstein-Hilbert gravity. Inclusion of matter in this theory has not yet been resolved. Nevertheless, it would be interesting to see the role of the complex monopoles in connection with such theories. Making contact between 3 dimensional gravity and the present work is the subject of a separate study \cite{BayramYutaka}. 

 Another interesting future direction is the connection with Josephson 
junction. The
three-dimensional compact QED is related to the Josephson 
junction system by an electro-magnetic dual transformation \cite{Hosotani} .
The $U(1)$ field strengths $(E_1, E_2, B)$ in the Georgi-Glashow model
correspond to $(B_1, B_2, E_3)$ in the Josephson junction.
Electric charges in the Georgi-Glashow model are magnetic charges
inserted in the barrier in the Josephson junction. 
At the moment we haven't understood what kind of an additional
interaction in the Josephson junction system corresponds 
to the Chern-Simons term in the Georgi-Glashow model.
It could be a $\theta F_{\mu\nu} \tilde F^{\mu\nu}$ term in the
superconductors on both sides.  Normally a 
$\theta F_{\mu\nu} \tilde F^{\mu\nu}$ term is irrelevant in QED.
However, if the values of $\theta$ on the
left and right sides are different, this term may result in a physical
consequence, which may mimic the effect of the Chern-Simons term in
the Georgi-Glashow model.

If monopoles are irrelevant in the presence of the Chern-Simons term,
it would imply that suppercurrents cease to flow across the barrier
in the corresponding Josephson junction.  Although we have not found
the precise analogue in the Josephson system yet, and therefore we cannot
say anything definite by analogy, we feel that it is very puzzling 
if suppercurrents suddenly stop to flow.  Monopoles should remain
important even in the presence of the Chern-Simons term. It is clear that further investigation is necessary for a better understanding of this subject.

\vspace{0.2in}

\noindent {\bf Acknowledgements}

\vspace{0.15in}

K.S. wishes to thank the organizers of Quarks-98 for the invitation, and for a stimulating  and well organized meeting in beautiful Suzdal. This work was supported in part  by the U.S. Department of Energy under contracts DE-FG02-94ER-40823.  

\vspace{0.2in}

\leftline{\bf References}  

\renewenvironment{thebibliography}[1]
        {\begin{list}{[$\,$\arabic{enumi}$\,$]}  % {\arabic{enumi}.}
        {\usecounter{enumi}\setlength{\parsep}{0pt}
         \setlength{\itemsep}{0pt}  \renewcommand{\baselinestretch}{1.2}
         \settowidth
        {\labelwidth}{#1 ~ ~}\sloppy}}{\end{list}}

\myend